\documentclass[preprint,aps]{revtex4-2}
\usepackage{mathrsfs}
\usepackage{graphicx}
\usepackage{multirow}
\usepackage{makecell}
\usepackage{booktabs}
\usepackage{color}
\usepackage{bm}

\begin{document}

\title{Spectroscopic Evidence for a Three-Dimensional Charge Density Wave in Kagome Superconductor CsV$_3$Sb$_5$}

\author{Cong Li$^{1,2,\sharp,*}$, Xianxin Wu$^{3,4,\sharp}$, Hongxiong Liu$^{5,\sharp}$, Craig Polley$^{6}$, Qinda Guo$^{1}$, Yang Wang$^{5}$, Xinloong Han$^{7}$, Maciej Dendzik$^{1}$, Magnus H. Berntsen$^{1}$, Balasubramanian Thiagarajan$^{6}$, Youguo Shi$^{5}$, Andreas P. Schnyder$^{3}$, Oscar Tjernberg$^{1,*}$
}

\affiliation{
\\$^{1}$Department of Applied Physics, KTH Royal Institute of Technology, Stockholm 11419, Sweden
\\$^{2}$Department of Applied Physics, Stanford University, Stanford, CA 94305, USA
\\$^{3}$Max-Planck-Institut f$\ddot{u}$rFestk$\ddot{o}$rperforschung, Heisenbergstrasse 1, D-70569 Stuttgart, Germany
\\$^{4}$CAS Key Laboratory of Theoretical Physics, Institute of Theoretical Physics, Chinese Academy of Sciences, Beijing 100190, China
\\$^{5}$Beijing National Laboratory for Condensed Matter Physics, Institute of Physics, Chinese Academy of Sciences, Beijing 100190, China
\\$^{6}$MAX IV Laboratory, Lund University, 22100 Lund, Sweden
\\$^{7}$Kavli Institute of Theoretical Sciences, University of Chinese Academy of Sciences, Beijing, 100049, China
\\$^{\sharp}$These people contributed equally to the present work.
\\$^{*}$Corresponding authors: conli@kth.se, oscar@kth.se
}

\pacs{}

\maketitle


{\bf The recently discovered AV$_3$Sb$_5$ (A=K, Rb, Cs) family, possessing V kagome nets, has received considerable attention due to the topological electronic structure and intriguing correlated phenomena, including an exotic charge density wave (CDW) and superconductivity. Detailed electronic structure studies are essential to unravel the characteristics and origin of the CDW as well as its interplay with superconductivity. Here, we present angle-resolved photoemission spectroscopy (ARPES) measurements for CsV$_3$Sb$_5$ at multiple temperatures and photon energies to reveal the nature of the CDW from an electronic structure perspective. We present evidence for a three-dimensional (3D) CDW order. In the process we also pinpoint a surface state attributed to a Cs terminated surface. This state was previously attributed to band folding band due to a CDW along the $c$ direction or a quantum well state from quantum confinement. The CDW expected 2-fold lattice reconstruction along $c$ axis is observed to be a quadrupling of the unit cell, thus for the first time directly demonstrating the 3D nature of the CDW from the electronic structure perspective. Moreover, this 3D CDW configuration originates from two distinct types of distortions in adjacent kagome layers. These present results not only provide key insights into the nature of the unconventional CDW in CsV$_3$Sb$_5$ but also provides an important reference for further studies on the relationship between the CDW and superconductivity.}


Traditionally, the kagome lattice provides an ideal playground to explore quantum magnetism\cite{LBalents_Nature2010,THHan_Nature2012_YSLee,YZhou_RMP2017_TKNg,CBroholm_Science2020_TSenthil}, such as the quantum spin liquid, owing to the intrinsic flat bands and geometric frustration. Recently, many novel quantum orders are discussed in the context of kagome systems\cite{WSWang_PRB2013_QHWang,SVIsakov_PRL2006_YBKim,MLKiesel_PRL2013_RThomale} and rich topological phenomena are discovered in transition-metal based kagome materials. These materials offer an exotic platform to study correlated quantum states intertwined with nontrivial band topology\cite{HMGuo_PRB2009_MFranz,JWen_PRB2010_GAFiete,HYang_NJP2017_BHYan,LDYe_Nature2018_JGCheckelsky,EKLiu_NP2018_CFelser,JXYin_NP2019_MZHasan,MGKang_NM2020_RComin,HMGuo_PRB2009_MFranz}. The recently discovered kagome metal family AV$_3$Sb$_5$ (A=K, Rb, Cs) exhibits a variety of intriguing phenomena, including the giant anomalous Hall effect\cite{SYYang_SA2020_MNAli,FHYu_PRB2021_XHChen} in the absence of magnetic order\cite{EMKenney_JPCM2021_MJGraf}, topologically nontrivial electronic structures\cite{BROrtiz_PRL2020_SDWilson,BROrtiz_PRM2021_SDWilson,MGKang_arxiv2021_RComin,YHu_arxiv2021_MShi_VHS,ZYHao_2021arxiv_CYChen}, an unconventional charge density wave (CDW)\cite{YXJiang_NM2020_MZHasan,ZWLiang_PRX2021_XHChen,HXTan_PRL2021_BHYan,ZHLiu_2021PRX_SCWang,YPLin_PRB2021_RMNandkishore,ZGWang_arxiv2021_ZXZhao,BROrtiz_PRX2021_SDWilson_CDW,JLuo_arxiv2021_GQZheng,KNakayama_PRB2021_TSato,HLi_arxiv2021_IZeljkovic,HSXu_PRL2021_DLFeng,YHu_arxiv2021_MShi,NShumiya_PRB2021_MZHasan,MGKang_arxiv2021_RComin,ZWWang_PRB2021_YGYao,YLuo_arxiv2021_JFHe,RLuo_arxiv2021_SBorisenko,HMiao_PRB2021_BHYan,HLLuo_arxiv2021_XJZhou,QWang_AM2021_YPQi,JWYu_arxiv2021_WLi} and a rare occurrence of superconductivity in a kagome system\cite{BROrtiz_PRL2020_SDWilson,BROrtiz_PRM2021_SDWilson,CCZhao_arxiv2021_SYLi,HChen_Nature2021_HJGao,QWYin_CPL2021_HCLei}. For this reason, the V-based kagome metals have attracted tremendous attention.

The CDW phase transition of AV$_3$Sb$_5$ occurs at 78$\sim$103K and its highly unconventional nature was derived from recent experiments\cite{BROrtiz_PRM2019_ESToberer}. A $2\times2$ in-plane distortion in the V kagome layers was observed in scanning tunnelling microscopy (STM) experiments and magnetic field dependent measurements further suggested a chiral CDW order\cite{BROrtiz_PRM2019_ESToberer,BROrtiz_PRL2020_SDWilson,YXJiang_NM2020_MZHasan}. Direct evidence of the time reversal symmetry breaking in the CDW phase is provided in recent muon spin relaxation ($\mu$SR) measurements\cite{CMielkeIII_arxiv2021_ZGuguchia}. It indicates that the CDW order may be intimately related to an anomalous Hall effect\cite{FHYu_PRB2021_XHChen} and unconventional superconductivity\cite{ZWLiang_PRX2021_XHChen,CCZhao_arxiv2021_SYLi,KYChen_PRL2021_JGCheng,HChen_Nature2021_HJGao,HSXu_PRL2021_DLFeng,FDu_PRB2021_HYuan,XXWu_PRL2021_RThomale}. Besides the in-plane distortion, the CDW order turns out to possess 3D character: a 2$\times$2$\times$2 lattice reconstruction is revealed by STM\cite{ZWLiang_PRX2021_XHChen}, hard X-ray scattering\cite{HXLi_PRX2021_HMiao} and nuclear magnetic resonance (NMR)\cite{DWSong_arxiv2021_XHChen} measurements and even a 2$\times$2$\times$4 lattice reconstruction is suggested in CsV$_3$Sb$_5$\cite{BROrtiz_PRX2021_SDWilson_CDW,JLuo_arxiv2021_GQZheng}. However, direct evidence of a 3D CDW order in the electronic structure have been missing so far. Exploring the electronic structure of the CDW order is not only necessary to understand its nature but also its origin, which is controversial so far\cite{HXLi_PRX2021_HMiao,XXZhou_2021PRB_HHWen,CZWang_arxiv2021_JHCho,TPark_2021PRB_LBalents,YPLin_2021PRB_RMNandkishore,HXTan_2021PRL_BHYan,NRatcliff_2021PRM_JWHarter,EUykur_2021arxiv_AATsirlin,YFXie_2021arxiv_PCDai}.

In this article, we present angle-resolved photoemission (ARPES) measurements and band structure calculations to investigate the effect of the CDW order on the electronic structure of CsV$_3$Sb$_5$. The surface states associated with a Cs covered surface are distinguished from what was previously believed to be quantum well states\cite{YQCai_arxiv2021_CYChen} or band folding effects due to a CDW order along the $c$ direction\cite{YLuo_arxiv2021_JFHe}. In addition to the 2-fold lattice reconstruction, we also observe a 4-fold lattice reconstruction along the $c$ direction. This CDW order is detected in CsV$_3$Sb$_5$ for the first time. The results provide key insights into the origin of the unconventional CDW in CsV$_3$Sb$_5$ and also provides a reference for further studies on the relationship between the CDW and superconductivity.

\begin{figure*}[tbp]
\begin{center}
\includegraphics[width=1\columnwidth,angle=0]{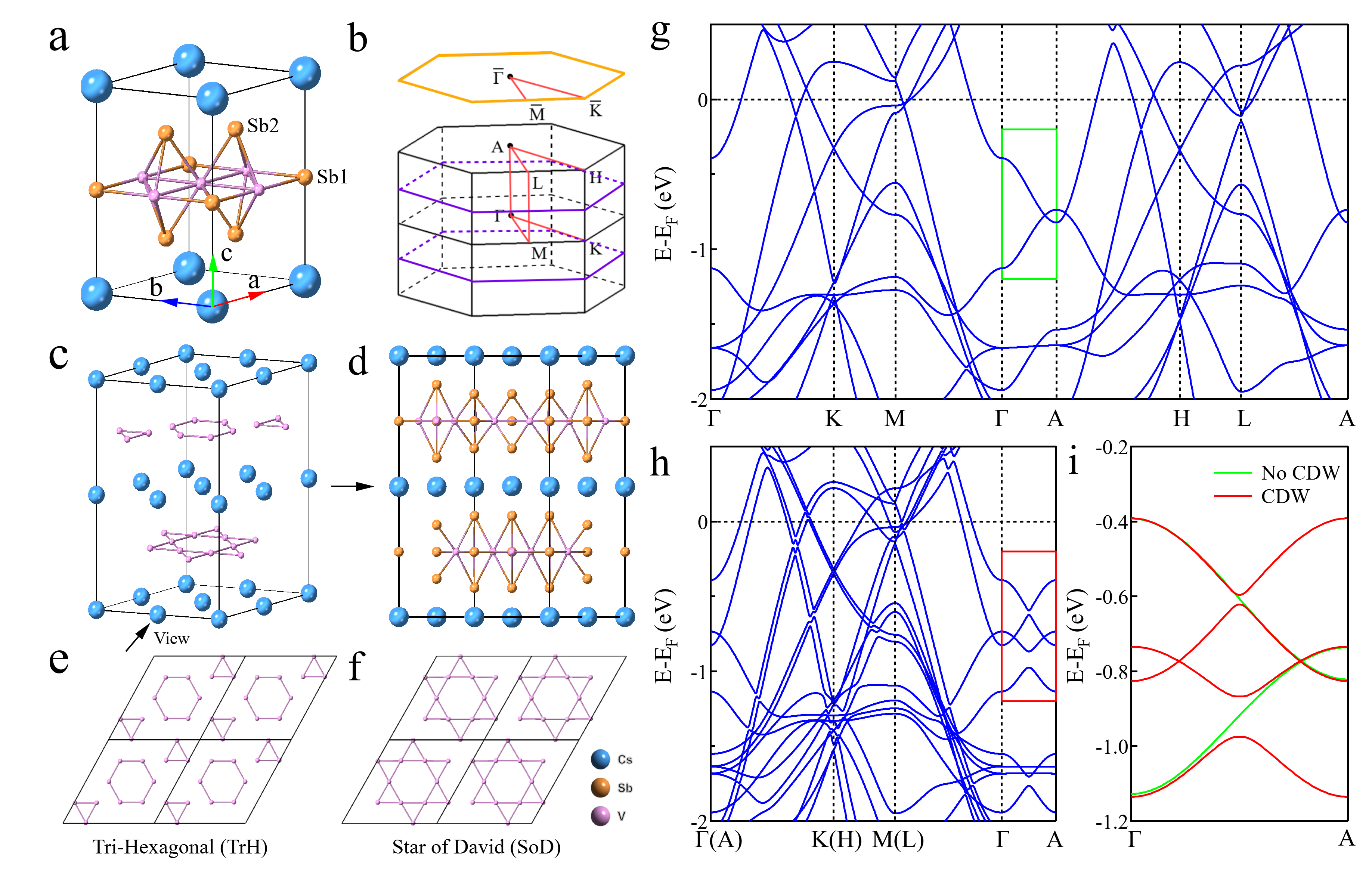}
\end{center}
\caption{\textbf{Crystal structure and band structure calculations of CsV$_3$Sb$_5$.} (a) The crystal structure of CsV$_3$Sb$_5$ with a space group P6/mmm. The Sb atom in the V-Sb plane is labeled as Sb1, and the Sb atom outside the V-Sb plane is labeled as Sb2. (b) Three-dimensional Brillouin zone (BZ) of the original unit cell (black lines) and considering a 2-fold reconstruction along $c$ axis (purple lines) of CsV$_3$Sb$_5$, and the corresponding two-dimensional BZ projected on the (001) plane (orange lines) in the pristine phase in (a). (c) One unit cell of CsV$_3$Sb$_5$ with considering the 2$\times$2$\times$2 lattice reconstruction due to the CDW. To get a clear view of the V atoms in the kagome layer, the Sb atoms are removed from it. (d) The side view of (c) but retains Sb atoms, where their distortion along $c$ axis is exaggerated. (e-f) Two different distortions of V atoms in the kagome layer due to the 2$\times$2$\times$2 lattice reconstruction caused by the CDW: one of them is tri-hexagonal (TrH) like distortion (e) and the other one is star of David (SoD) like distortion (f). (g-h) Calculated band structure of CsV$_3$Sb$_5$ along high symmetry directions across the Brillouin zone without (g) and with (h) considering the 2-fold lattice reconstruction along $c$ axis due to the CDW. (i) The zoom in of the band structure along $\Gamma$-A direction with (red curves) and without (green curves) considering the 2-fold lattice reconstruction along $c$ axis due to the CDW.
}
\label{fig1}
\end{figure*}

CsV$_3$Sb$_5$ has a layered crystal structure with the space group P6/mmm (no.191). The vanadium atoms form a kagome lattice that consists of a two-dimensional network of corner-sharing triangles, which are intercalated by Sb atoms forming a honeycomb lattice. The V$_3$Sb$_5$ kagome planes are separated by layers of Cs ions forming triangular networks (Fig. 1a). Below the CDW transition temperature (T$_{CDW}$=94~K) of CsV$_3$Sb$_5$, the Vanadium kagome planes exhibit a 2$\times$2 reconstruction suggested by STM measurements\cite{HZhao_Nature2021_IZeljkovic,HSXu_PRL2021_DLFeng,HChen_Nature2021_HJGao}. There are two possible distortions: a tri-hexagonal (TrH) distortion (Fig. 1e) or a star of David (SoD) distortion (Fig. 1f). SoD-/TrH-like distortions in the kagome layer will further induce a postive/negative out-of-plane A$_{1g}$ distortion mode of Sb2 atoms, as shown in Fig. 1d, implying an intrinsic three-dimensional (3D) feature of the CDW order. As a matter of fact, recent experiments suggest a 3D 2$\times$2$\times$2 lattice reconstruction\cite{ZWLiang_PRX2021_XHChen,HXLi_PRX2021_HMiao,DWSong_arxiv2021_XHChen}. Although the details of this 3D reconstruction are still lacking, it may be related to the distinct distortions in the adjacent kagome layers. Fig.~\ref{fig1}c displays one possible configuration with alternating stacking of SoD- and TrH- distortions. To illustrate the effect of CDW order on the electronic structure, we first show the normal electronic structure in Fig.~\ref{fig1}g from density functional theory (DFT) calculations (the orbital-resolved band structure is shown in Fig. S1 in the Supplemental Material). In the 2$\times$2$\times$2 CDW order phase, there are in-plane and out-of-plane folding and the folded band is displayed in Fig. S2 in the Supplemental Material. Here, we focus on the folding solely along the $c$ axis. When the lattice is doubled along $c$ axis in CsV$_3$Sb$_5$, which is introduced artificially in our calculations by moving the Cs atoms along $c$ axis, the band structure in the $k_z=\pi/c$ plane will be folded onto the $k_z=0$ plane, as shown in Fig.~\ref{fig1}h. Moreover, by comparing the band structure along the $\Gamma$-A direction in Fig.~\ref{fig1}i, with CDW order (red curves) and without (green curves), we can identify band doubling and the folding gap at $k_z=\pi/2c$, with the latter being a smoking-gun evidence for a 3D CDW order.


Figure 2 displays the comparison between ARPES measurements and DFT calculated electronic structure of CsV$_3$Sb$_5$. Fig. 2a shows the Fermi surface map of CsV$_3$Sb$_5$ measured at 150~K with a photon energy of 57~eV. A circular pocket centered at the BZ center can be identified as well as two triangular hole-like pockets around $\overline{K}$. By comparing with DFT calculations (Fig. S1 in Supplemental Material), we find that the circular Fermi pocket around the BZ center is mainly dominated by the $p_z$ orbital of the Sb atoms. Fig. 2b shows the calculated Fermi surface at $k_z$=$\pi/c$ for the pristine crystal structure in Fig. 1a, which is consistent with the Fermi surface observed in Fig. 2a. Fig. 2c-2d show the measured band structure along the $\overline{K}$-$\overline{\Gamma}$-$\overline{K}$ direction for temperatures above (150 K, Fig. 2c) and below (20 K, Fig. 2d) the CDW transition temperature (T$_{CDW}$=94 K). Both measurements are performed on freshly cleaved surfaces. However, no folding of the band structure (Fig. 2d) or Fermi surface reconstruction (Fig. S3 in the Supplemental Material) due to the 2$\times$2 in-plane CDW at 20 K is discernible. This is despite the fact that this reconstruction was observed in several STM measurements\cite{ZWLiang_PRX2021_XHChen,HZhao_Nature2021_IZeljkovic,HSXu_PRL2021_DLFeng,HChen_Nature2021_HJGao}. To date, no ARPES measurement has reported on the observation of in-plane CDW folding in CsV$_3$Sb$_5$. Only in KV$_3$Sb$_5$ has this been reported\cite{HLLuo_arxiv2021_XJZhou}. This may suggest a weaker in-plane CDW folding in CsV$_3$Sb$_5$. Further comparing Fig. 2c and 2d, it can be found that only one electron-like band ($\alpha$ band in Fig. 2c) can be observed in the vicinity of the BZ center at 150 K (Fig. 2c), but two electron-like bands ($\alpha$ and $\gamma$ bands in Fig. 2d) can be observed at 20 K (Fig. 2d). In addition, it is noted that when the temperature is lowered from 150 K to 20 K, the $\beta$ band splits into two bands. To understand our observations, we performed DFT band structure calculations without (Fig. 2e) and with (Fig. 2f) the 2-fold lattice reconstruction along $c$ axis due to the CDW. It can be noted that all bands in Fig. 2c can be well explained by Fig. 2e. When the temperature drops below the CDW transition temperature, the splitting of the $\beta$ band in Fig. 2d can be interpreted as a band folding due to the CDW along the $c$ axis (Fig. 2f) according to our calculation. Moreover, we see that the $\gamma$ band around the BZ center in Fig. 2d is similar in energy position to the folded band induced by the CDW in Fig. 2f. This band has been observed by several other groups using various photon energies\cite{YLuo_arxiv2021_JFHe,YQCai_arxiv2021_CYChen}, but its origin is still under debate. There are mainly two explanations: CDW folding along the $c$ axis\cite{YLuo_arxiv2021_JFHe}, or quantum well states due to quantum confinement\cite{YQCai_arxiv2021_CYChen}.


To further reveal the origin of the $\gamma$ band in Fig. 2d, we performed detailed photon energy dependent measurements. The data is shown in Fig. 3. Firstly, we examine the band structure close to the $\bar{\Gamma}$-point along the $\overline{K}$-$\overline{\Gamma}$-$\overline{K}$ direction for the sample cleaved at low temperature (20 K), as shown in the top panel of Fig. 3. Fig. 3a shows the photon energy dependent ARPES spectral intensity map at the binding energy of 0.2 eV along the $\overline{K}$-$\overline{\Gamma}$-$\overline{K}$ direction measured at 20 K. The corresponding photon energy dependent band structures is shown in Fig. 3b. Partially observable bands are labelled by red, green and orange dashed lines in Fig. 3b. The corresponding band minima from Fig. 3b are summarized in Fig. 3c, where one prominent feature is that the red band has almost no dispersion along $k_z$. The bands' momentum positions at a binding energy of 0.2~eV are extracted from Fig. 3b and displayed in Fig. 3a with red, green and orange open circles. Fig. 3d-3f show similar data as Fig. 3a-3c but at a temperature of 120 K. At this temperature, it is evident that only one band in the vicinity of the $\overline{\Gamma}$ point can be observed and the red dashed band in Fig. 3b is absent in Fig. 3e. This also becomes clear by comparing the photon energy dependent ARPES spectral intensity maps for the two temperatures shown in Fig. 3a and Fig. 3d. Furthermore, the bottom of the band marked in Fig. 3e experiences a photon-energy dependence and reaches an energy minimum at a photon energy around 49 eV. The temperature dependence of the bands around $\bar{\Gamma}$ in samples cleaved at low temperature further motivates us to carry out additional measurements at low temperature for the sample cleaved at high temperature (150~K). The data is shown in Fig. 3g-3i. By comparing Fig. 3a-3b and Fig. 3g-3h we are looking at data collected at the same temperature (20 K) but for samples with different cleave temperatures. It is clear that the red dashed band around the $\overline{\Gamma}$ point can only observed in the sample cleaved at low temperature (20 K) but not in the sample cleaved at high temperature (150 K). This clearly indicates that the $\gamma$ band in Fig. 2d does not originate from the CDW order. In addition, the $\gamma$ band in Fig. 2d also cannot be attributed to quantum well states from quantum confinement, as they should be observed in all measurements independent of cleave and measurement temperature. We believe that the $\gamma$ band is most likely a surface state due to its non-dispersive nature along the $k_z$ direction (red open circle in Fig. 3a). To confirm this, we performed theoretical calculations for a slab of CsV$_3$Sb$_5$ with two distinct terminations (Cs or Sb) and the band structure can be found in Fig. S4 in the Supplemental Material. The surface state in blue (Fig. S4b in the Supplemental Material) around $\overline{\Gamma}$ is mainly attributed to surface Sb1 atoms in the Cs-terminated surface. It is below the bulk bands in energy owing to the electron doping from the top Cs surface layer. This surface is consistent with the $\gamma$ band in Fig. 2d and Fig. 3b in our measurement. Many STM measurements show that there are two types of terminated surfaces in CsV$_3$Sb$_5$: a hexagonal lattice that is attributed to the Cs layer, and a honeycomb-like surface structure ascribed to the Sb layer\cite{HZhao_Nature2021_IZeljkovic,HSXu_PRL2021_DLFeng,HChen_Nature2021_HJGao}. However, the Cs surface layer is unstable, often resulting in randomly distributed Cs atoms prone to clustering\cite{HZhao_Nature2021_IZeljkovic}, and they can escape from the surface at high temperatures. This process is irreversible and leads to dominant Sb terminated surface at high temperature. Therefore, the surface states of Cs-terminated surfaces are usually not observed neither at high measurement temperatures for samples cleaved at a low temperature nor at low measurement temperatures for samples cleaved at a high temperature. From the above three measurements (Fig. 3b, 3e and 3h), we can infer that the bands labelled by the green dashed line are the original bands and the bands labelled by the orange dashed line are directly related to the CDW order. We also notice that the bands caused by the CDW folding in Fig. 3h-3i exhibit an energy shift towards the Fermi level relative to the CDW folded bands in Fig. 3b-3c, which may be attributed to hole doping from the escape or oxidation of surface Cs atoms at high temperature\cite{YPSong_PRL2021_XLChen}.


After having identified the surface state, we study the band folding along $k_z$ in the CDW ordered state. According to the above analysis, the photon energy near 49 eV in Fig. 3e should correspond to the A point in the BZ, from which the inner potential V$_0$ of CsV$_3$Sb$_5$ is estimated to be approximately 7.3 eV. We then convert the photon energy dependent ARPES spectral intensity map (Fig. S5 in Supplemental Material) into a Fermi surface map along the $k_z$ direction. Figs. 4a-4d show constant energy maps around the BZ center along the $k_z$ and $\overline{K}$-$\overline{\Gamma}$-$\overline{K}$ directions at binding energies of 0, 0.2, 0.4 and 0.5 eV, respectively. It is evident that these contours show a periodic variation along the $k_z$ direction. Fig. 4e displays the band structure along the $k_z$ direction close to the BZ center (the corresponding photon energy dependent spectra for photoemission intensity at the BZ center are shown in Fig. S6 in the Supplemental Material). Due to matrix element effects and variations in the beam intensity at different photon energies, we perform intensity normalization of the measured data  based on the intensity in the energy range of -0.8 to -0.5 eV to make the EDC intensities at different photon energies uniform. Fig. 4f shows stacked EDCs extracted from the original band structure along the $k_z$ direction without normalization. The green open circles mark the peak position of main subbands in Fig. 4e which are extracted from Fig. 4f. Fig. 4g shows the corresponding EDC second derivative image of Fig. 4e, with renormalization around -0.8 eV. The extracted peak positions of main subbands of Fig. 4f are replotted in Fig. 4h together with bands from the theoretical calculations. The blue solid curves are the calculated band structures along the $k_z$ direction close to the BZ center with a  doubling of the unit cell along $c$, simulating the $2\times2\times2$ CDW order. The observed band folding around $k_z$=8.5, 11, 12.5 $\pi/c$, labelled by blue arrows, is consistent with calculations. Moreover, the calculated gap resulting from the folding at $k_z=n\pi/2c$ ($n$ is an integer) is also directly observed at $k_z=8.5$ and $12.5$ $\pi/c$, although it is relatively small in the experiment. However, the bands marked by red arrows in Fig. 4h, especially the appearance of folding at $k_z=n\pi/c$, cannot be explained by the folding from a doubling of the unit cell along the $c$ axis. Their $\pi/2c$ shift relative to the blue curves further motivates us to consider a quadrupling of the unit cell along $c$ in the CDW phase and the corresponding bands due to this 4-fold folding are shown as dashed red curves in Fig. 4h. In this case, the bands at $k_z=10.5$, $13.5$, $14$ and $14.5$ $\pi/c$ are consistent with calculations. Thus, the experimental bands  along $k_z$ close to the BZ center (green open cirlces in Fig. 4h) can be well explained (red dashed lines in Fig. 4h) within the range of experimental error. The observed band dispersion from our ARPES measurements is the first direct evidence for the 3D nature of a CDW order with a quadrupling of the unit cell along the $c$ axis in CsV$_3$Sb$_5$. Our observation of the 4-fold lattice reconstruction along the $c$ axis due to the CDW is consistent with previous X-ray diffraction\cite{BROrtiz_PRX2021_SDWilson_CDW} and nuclear magnetic resonance (NMR)\cite{JLuo_arxiv2021_GQZheng} measurements.\\


The above observation of out-of-plane band folding directly demonstrates the 3D nature of the CDW order. The SoD-/TrH-like distortions in the V kagome layer can induce a postive/negative out-of-plane $A_{1g}$ distortion mode on Sb2 atoms, which naturally introduces interlayer coupling between distorted kagome layers. These distortions further couple with Sb1 $p_z$ orbitals along $c$ axis, resulting in the observed band folding along the $\Gamma$-$A$ direction. Moreover, the observed small folding gap indicates that the distortion along $c$ axis in the CDW phase couples weakly with Sb1 $p_z$ orbital and is related to different stacking of distortions in V kagome layers.

With permutations of SoD-like and TrH-like distortions of kagome layer along the $c$ direction, two kinds of stacking can occur. The first one is the alternate stacking of the SoD-like and TrH-like distortions along the $c$ axis (Case-1), leading to a $2\times2\times2$ CDW order, as shown in Fig. S2a-S2b in Supplemental Material. The other case is the alternating stacking of the SoD- (TrH-) like distortions with an in-plane $\pi$ phase shift (Case-2), as shown in Fig. S2c (TrH-like) and Fig. S2d (SoD-like) in Supplemental Material~\cite{DWSong_arxiv2021_XHChen}. However, different stacking will lead to distinctly folded bands along $c$ axis. In contrast to the same out-of-plane distortion for each layer in the Case-2, the opposite out-plane distortions between adjacent layers in the Case-1 induce stronger charge modulation along $c$ axis, leading to a stronger $k_z$ folding. It is also confirmed by theoretical calculations that the folding gaps of Sb $p_z$ band at $k_z=\pi/2c$ are larger in the Case-1. Moreover, the band structures in these two cases exhibit different behaviour around the M point in the BZ (Fig. S7 in Supplemental Material). For the Case-1, the $\delta$ band around the M point split into two bands ($\delta_1$ and $\delta_2$ bands) (Fig. S7a-S7b) due to the CDW order, but for the Case-2, the $\delta$ band does not split (Fig. S7c-S7d). Based on our ARPES measurements (Fig. S7e-f) and the observed prominent $k_z$ folding (Fig. 4h), it can be inferred that the 3D CDW order may originate from the alternating stacking of the SoD- and TrH-like distortions along the $c$ axis. Furthermore, a `TTSS' (T: TrH, S: SoD) or `T$\overline{\mathrm{T}}$S$\overline{\mathrm{S}}$' (`T$\overline{\mathrm{T}}$$\overline{\mathrm{S}}$S', `$\overline{\mathrm{T}}$TS$\overline{\mathrm{S}}$', `$\overline{\mathrm{T}}$T$\overline{\mathrm{S}}$S', `T$\overline{\mathrm{S}}$$\overline{\mathrm{T}}$S', `TS$\overline{\mathrm{T}}$$\overline{\mathrm{S}}$') ($\overline{\mathrm{T}}$: TrH with $\pi$ phase in-plane shift, $\overline{\mathrm{S}}$: SoD with $\pi$ phase in-plane shift) stacking will result in a quadrupled unit cell along the $c$ axis, which can induce a quadruple folding at $k_z=n\pi/c, n\pi/2c$ ($n$ is an integer) as observed in our measurements. Compared with the out-of-plane direction, it is difficult to detect any noticeable in-plane folding in our measurements, despite the $2\times2$ in-plane pattern revealed in STM measurements\cite{ZWLiang_PRX2021_XHChen,HZhao_Nature2021_IZeljkovic,HSXu_PRL2021_DLFeng,HChen_Nature2021_HJGao}. Supporting evidence of in-plane folding has, on the other hand, been provided in KV$_3$Sb$_5$ by laser ARPES measurements\cite{HLLuo_arxiv2021_XJZhou}. The difference between the two systems may be related to the complicated stacking of SoD- and TrH-like distortions along the $c$ axis in CsV$_3$Sb$_5$, which could generate destructive in-plane folding.

The observed CDW order can be suppressed by external pressure and vanishes around 2~GPa, at which point the superconducting transition temperature reaches its maximum\cite{KYChen_PRL2021_JGCheng}. In this regime, the in-plane lattice constant $a$ shows negligible change while the out-of-plane lattice constant $c$ is significantly reduced\cite{KYChen_PRL2021_JGCheng}. Moreover, recent experiments show that the CDW order is suppressed in thin films\cite{YPSong_PRL2021_XLChen,BQSong_arxiv2021_SYLi}. These observations imply that moderate interlayer coupling stabilizes the CDW order but strong interlayer coupling suppresses it, demonstrating that the 3D CDW order can be tuned through the coupling strength along the $c$ axis. As the CDW order competes with superconductivity, studying the origin of CDW order can be helpful for understanding the mechanism behind superconductivity. To further understand the primary driving force for the CDW order, studying thinner films or even monolayer kagome metals would become crucial. So far the origin of the CDW order is still controversial\cite{HXLi_PRX2021_HMiao,XXZhou_2021PRB_HHWen,CZWang_arxiv2021_JHCho,TPark_2021PRB_LBalents,YPLin_2021PRB_RMNandkishore,HXTan_2021PRL_BHYan,NRatcliff_2021PRM_JWHarter,EUykur_2021arxiv_AATsirlin,YFXie_2021arxiv_PCDai}. Our spectroscopic observation of the 3D character provides crucial insights into the nature of the CDW order. Particularly, the observation of quadruple folding along the $c$ axis, if it is a general feature of all AV$_3$Sb$_5$ kagome metals,  would support the importance of electron-phonon interaction in promoting the CDW order as Fermi surface nesting at the corresponding vector is not prominent.


In summary, we performed angle-resolved photoemission spectroscopy(ARPES) measurements and band structure calculations to investigate the electronic structure of CsV$_3$Sb$_5$. Band features that were previously interpreted as band folding due to a CDW or quantum well states are demonstrated to be surface states associated with the Cs terminated surface. In addition to the 2-fold lattice reconstruction, a 4-fold lattice reconstruction along the $c$ direction, driven by the CDW order, is observed for the first time in electronic structure measurements on CsV$_3$Sb$_5$. These results provide key insights to the origin of the unconventional CDW in CsV$_3$Sb$_5$ and also provides a reference for further studies on the relationship between the CDW and superconductivity.


\noindent {\bf Methods}\\
\noindent{\bf Sample} Single crystals of CsV$_3$Sb$_5$ were grown from CsSb$_2$ alloy and Sb as flux. Cs, V, Sb elements and CsSb$_2$ precursors were sealed in a Ta crucible in a molar ratio of 1:3:14:10, which was finally sealed in a highly evacuated quartz tube. The tube was heated up to 1273 K, dwelt for 20 hours and then cooled down to 763 K slowly. Single crystals with silvery luster were separated from the flux by centrifuging.

\noindent{\bf ARPES Measurements} High-resolution ARPES measurements were performed at the Bloch beamline of MAX IV. The total energy resolution (analyzer and beamline) was set at 15 meV for the measurements. The angular resolution of the analyser was $\sim$0.1 degree. The beamline spot size on the sample was about 12 $\mu$m$\times$15 $\mu$m. All measurements were carried out with linear-horizontal (LH) polarization. The samples were cleaved {\it in situ} and measured at different temperatures in ultrahigh vacuum with a base pressure better than 1.0$\times$10$^{-10}$ mbar.

\noindent{\bf DFT calculations} Our Density functional theory (DFT) calculations employ the projector augmented wave method encoded in Vienna ab initio simulation package, and the local density approximation for the exchange correlation functional is used. Throughout this work, the cutoff energy of 500 eV is taken for expanding the wave functions into plane-wave basis. In the calculation, the Brillouin zone is sampled in the $k$ space within Monkhorst-Pack scheme. The number of these k points depends on unitcells: 10$\times$10$\times$6 for normal unit cell, 10$\times$10$\times$3 for $1\times1\times2$ super cell and 5$\times$5$\times$3 for $2\times2\times2$ super cell. We used the experimental lattice parameters $a=5.495$\AA~ and $c=9.309$\AA~ for the normal unit cell. In the $2\times2\times2$ supercell, the alternating stacking of SoD- and TrH- distortion in V kagome layers is initialized and then we performed the relaxation of atomic positions. To simulate the band folding along $c$ axis from the 3D CDW order but avoid complex in-plane folding, we introduce Cs movement along the $c$ axis while the V kagome lattice remains undistorted.

\vspace{3mm}

\noindent {\bf Acknowledgement}\\
The work presented here was financially supported by the Swedish Research council (2019-00701) and the Knut and Alice Wallenberg foundation (2018.0104). M.D. acknowledges financial support from the Göran Gustafsson Foundation. Q.D.G acknowledges the fellowship from Chinese scholarship council (No.201907930007). Y.G.S. acknowledges the National Natural Science Foundation of China (Grants No. U2032204), and the K. C. Wong Education Foundation (GJTD-2018-01).

\vspace{3mm}

\noindent {\bf Author Contributions}\\
C.L., X.X.W. and H.X.L. contributed equally to this work. C.L. proposed and designed the research. H.X.L. and Y.G.S. contributed to CsV$_3$Sb$_5$ crystal growth. X.X.W., X.L.H and A.P.S. contributed to the band structure calculations and theoretical discussion. C.L. carried out the experiment with the assistance from C.P. and Q.D.G.. C.L. contributed to software development for data analysis. C.L. analyzed the data. C.L. wrote the paper with X.X.W., A.P.S. and O.T.. All authors participated in discussion and comment on the paper.

\newpage

\begin{figure*}[tbp]
\begin{center}
\includegraphics[width=1\columnwidth,angle=0]{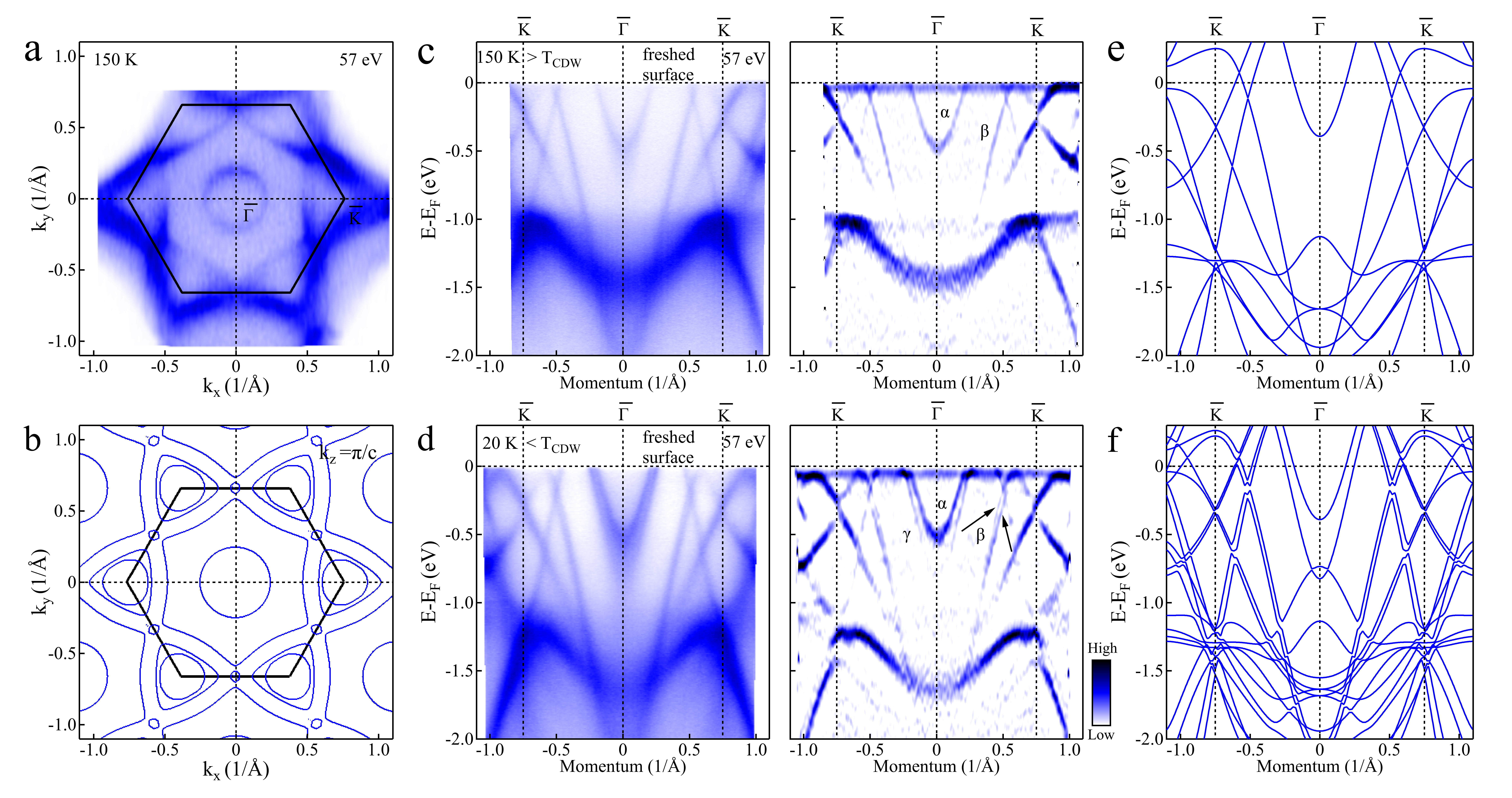}
\end{center}
\caption{\textbf{Fermi surface mapping and band structures of CsV$_3$Sb$_5$.} (a) The Fermi surface mapping of CsV$_3$Sb$_5$ measured at 150 K with a photon energy of 57 eV. (b) Calculated Fermi surface of CsV$_3$Sb$_5$ at $k_z$=$\pi/c$ plane. The BZ is defined by black solid lines. (c-d) The band structures (left side) and the corresponding energy distribution curve (EDC) second derivative image along $\overline{K}$-$\overline{\Gamma}$-$\overline{K}$ direction measured at 150 K (c) and 20 K (d) with the photon energy of 57 eV. The sample in (c) is cleave at 150 K, and the sample in (d) is cleave at 20 K. (e-f) Calculated band structure of CsV$_3$Sb$_5$ along $\overline{K}$-$\overline{\Gamma}$-$\overline{K}$ direction without (e) and with (f) considering the 2-fold lattice reconstruction along $c$ axis due to the CDW.
}
\label{fig2}
\end{figure*}

\begin{figure*}[tbp]
\begin{center}
\includegraphics[width=1\columnwidth,angle=0]{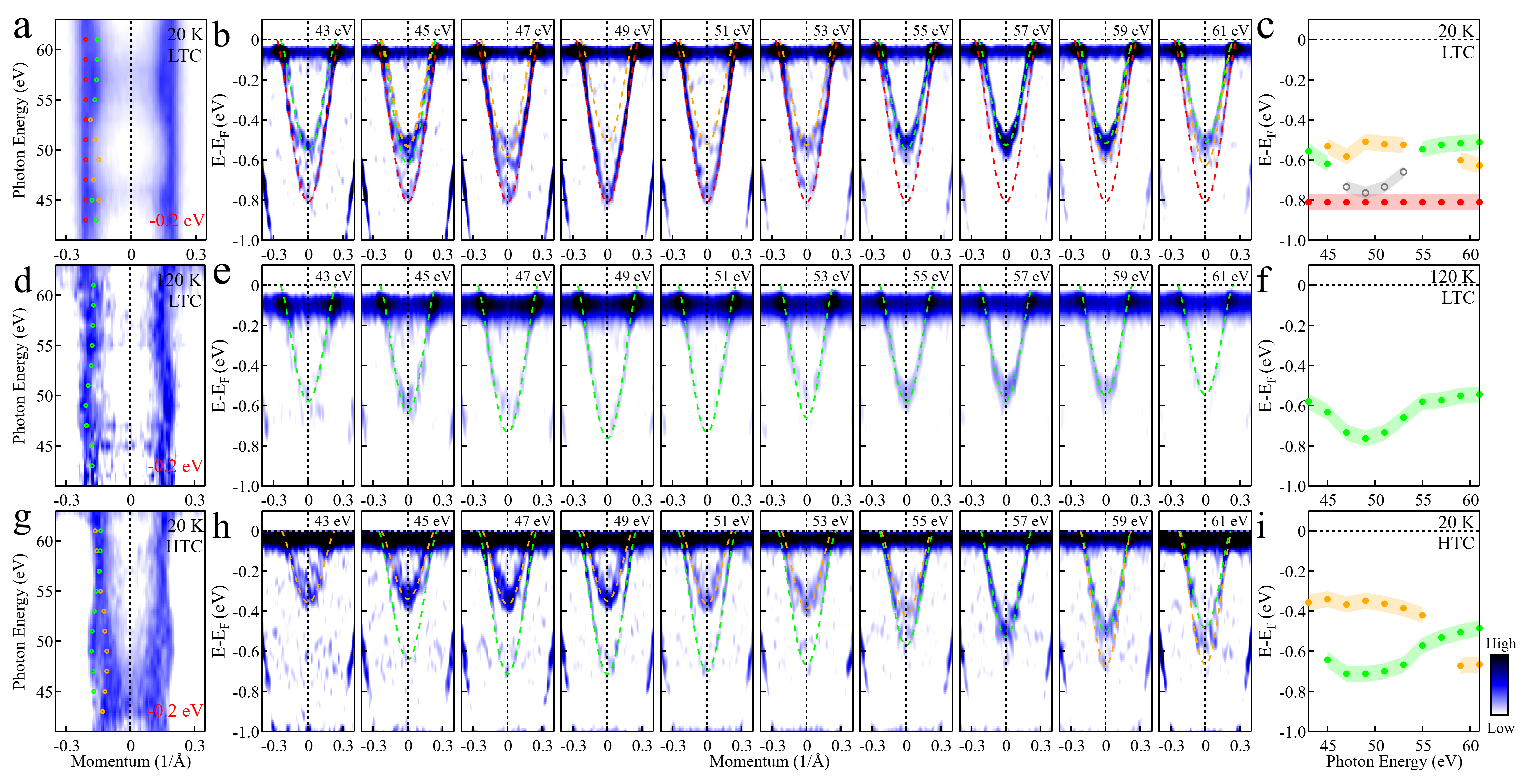}
\end{center}
\caption{\textbf{Photon energy dependent electronic structures of CsV$_3$Sb$_5$.} (a) Photon energy dependent ARPES spectral intensity map at the binding energy of 0.2 eV along the $\overline{K}$-$\overline{\Gamma}$-$\overline{K}$ direction measured at 20 K with the samples of low temperature (20 K) cleavage (LTC). (b) Band structure around the BZ center measured with a series of photon energies from 43 eV to 61 eV. Main subbands are labelled by the red (surface states), green (original bands) and orange (CDW folding bands) dashed lines (guide to the eye). (c) The extracted band bottom from the main subbands in (b). (d-f) The similar measurements to (a-c), but acquired at a sample temperature of 120 K. (g-i) The similar measurements to (a-c), but with the sample cleaved at 150 K and subsequently cooled to 20 K for the measurements. For comparison, the data points corresponding to photon energies ranging from 47 eV to 53 eV in (f) are re-plotted in (c) with the grey open circle. The red, green and orange open circle in (a), (d) and (g) are the momentum positions for the bands extracted from panels (b), (e) and (h), respectively, at a binding energy of 0.2 eV.
}
\label{fig3}
\end{figure*}

\begin{figure*}[tbp]
\begin{center}
\includegraphics[width=1\columnwidth,angle=0]{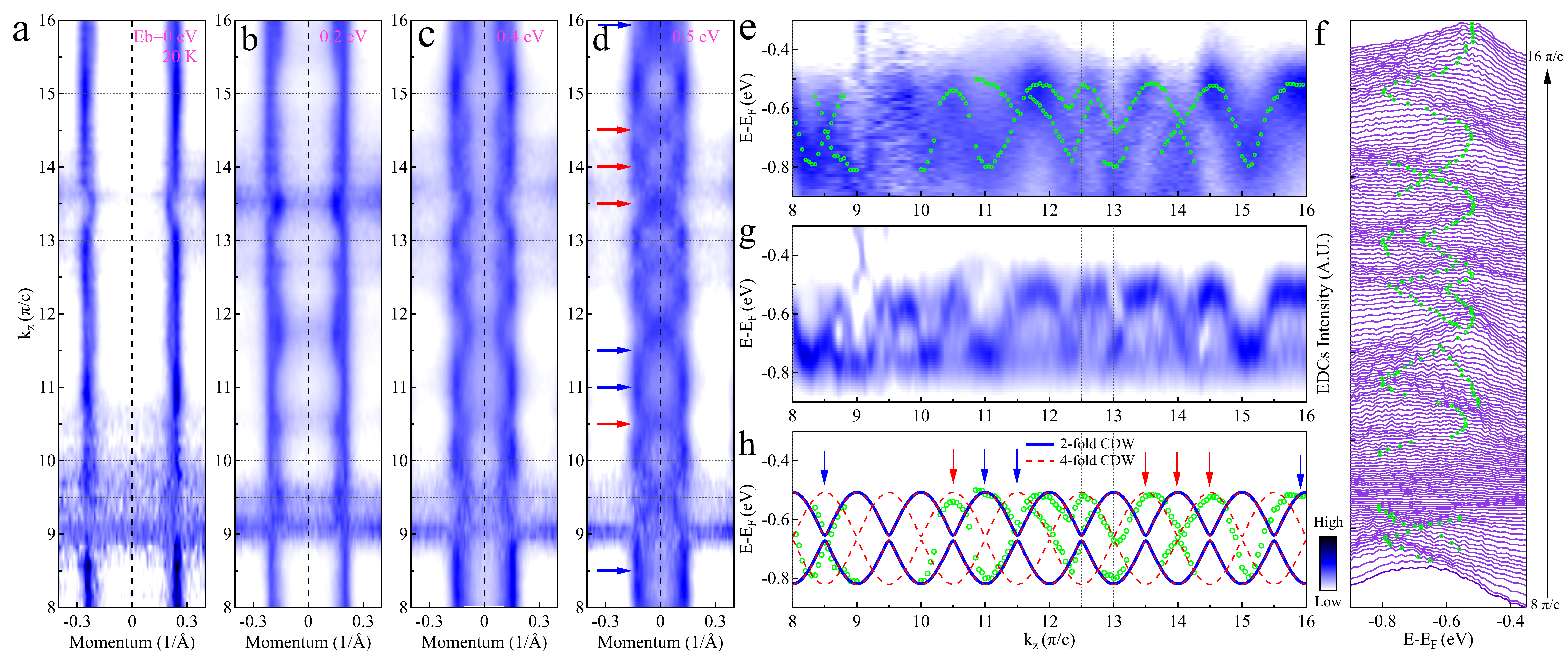}
\end{center}
\caption{\textbf{Electronic structure folding due to the CDW along the $c$ axis.} (a-d) Photon energy dependent measurements of CsV$_3$Sb$_5$ crossing the zone center along $\overline{K}$-$\overline{\Gamma}$-$\overline{K}$ direction at a binding energy of 0 eV (a), 0.2 eV (b), 0.4 eV (c) and 0.5 eV (d). (e) The band structure along the $k_z$ direction around the BZ center. The EDC intensities are normalized in the range of -0.8 eV to -0.5 eV. The green open circles mark the peak position of the main subbands which are extracted from (f) by fitting EDCs with Lorentzian curves combined with manually determining the position of local extrema in some regions that were fitting was difficult. (f) The EDCs stack of (e) but without intensity normalized. (g) The corresponding EDC second derivative image of (e) normalized around -0.8 eV. For comparison purposes, the extracted peak position of main subbands of (f) are replotted in (h). The blue solid (red dashed) curves are the calculated band structures along the $k_z$ direction around the BZ center when considering 2-fold (4-fold) lattice reconstruction along $c$ axis due to the CDW. In order to reach a quantitative agreement with the experimental data, we have rescaled the calculated band structures in energy with a factor 0.72 and introduced an energy shift of -0.225 eV. The blue (red) arrows in (h) are mark the band features that can be explained by blue (red dashed) lines. The corresponding arrows are also labelled in (d).
}
\label{fig4}
\end{figure*}


\begin{thebibliography}{99}


\bibitem{LBalents_Nature2010} L. Balents, Spin liquids in frustrated magnets. Nature {\bf 464}, 199 (2010).

\bibitem{THHan_Nature2012_YSLee} T. H. Han \emph{et al}., Fractionalized excitations in the spin-liquid state of a kagome-lattice antiferromagnet. Nature {\bf 492}, 402 (2012).

\bibitem{YZhou_RMP2017_TKNg} Y. Zhou \emph{et al}., Quantum spin liquid states. Rev. Mod. Phys. {\bf 89}, 025003 (2017).

\bibitem{CBroholm_Science2020_TSenthil} C. Broholm \emph{et al}., Quantum spin liquids. Science {\bf 367}, 263 (2020).


\bibitem{WSWang_PRB2013_QHWang} W. S. Wang \emph{et al}., Competing electronic orders on kagome lattices at van Hove filling. Phys. Rev. B {\bf 87}, 115135 (2013).

\bibitem{SVIsakov_PRL2006_YBKim} S. V. Isakov \emph{et al}., Hard-Core Bosons on the Kagome Lattice: Valence-Bond Solids and Their Quantum Melting. Phys. Rev. Lett. {\bf 97}, 147202 (2006).

\bibitem{MLKiesel_PRL2013_RThomale} M. L. Kiesel \emph{et al}., Unconventional Fermi Surface Instabilities in the Kagome Hubbard Model. Phys. Rev. Lett. {\bf 110}, 126405 (2013).


\bibitem{HMGuo_PRB2009_MFranz} H. M. Guo \emph{et al}., Topological insulator on the kagome lattice. Phys. Rev. B {\bf 80}, 113102 (2009).

\bibitem{JWen_PRB2010_GAFiete} J. Wen \emph{et al}., Interaction-driven topological insulators on the kagome and the decorated honeycomb lattices. Phys. Rev. B {\bf 82}, 075125 (2010).

\bibitem{HYang_NJP2017_BHYan} H. Yang \emph{et al}., Topological Weyl semimetals in the chiral antiferromagnetic materials Mn$_3$Ge and Mn$_3$Sn. New J. Phys. {\bf 19}, 015008 (2017).

\bibitem{LDYe_Nature2018_JGCheckelsky} L. D. Ye \emph{et al}., Massive Dirac fermions in a ferromagnetic kagome metal. Nature {\bf 555}, 638 (2018).

\bibitem{EKLiu_NP2018_CFelser} E. K. Liu \emph{et al}., Giant anomalous Hall effect in a ferromagnetic kagome-lattice semimetal. Nat. Phys. {\bf 14}, 1125 (2018).

\bibitem{JXYin_NP2019_MZHasan} J. X. Yin \emph{et al}., Negative flat band magnetism in a spin–orbit-coupled correlated kagome magnet. Nat. Phys. {\bf 15}, 443 (2019).

\bibitem{MGKang_NM2020_RComin} M. G. Kang \emph{et al}., Dirac fermions and flat bands in the ideal kagome metal FeSn. Nat. Mater. {\bf 19}, 163 (2020).



\bibitem{SYYang_SA2020_MNAli} S. Y. Yang \emph{et al}., Giant, unconventional anomalous Hall effect in the metallic frustrated magnet candidate, KV$_3$Sb$_5$. Sci. Adv. {\bf 6}, eabb6003 (2020).

\bibitem{FHYu_PRB2021_XHChen} F. H. Yu \emph{et al}., Concurrence of anomalous Hall effect and charge density wave in a superconducting topological kagome metal. Phys. Rev. B {\bf 104}, L041103 (2021).

\bibitem{EMKenney_JPCM2021_MJGraf} E. M. Kenney \emph{et al}., Absence of local moments in the kagome metal KV$_3$Sb$_5$ as determined by muon spin spectroscopy. J. Phys. Condens. Matter {\bf 33}, 235801 (2021).


\bibitem{BROrtiz_PRL2020_SDWilson} B. R. Ortiz \emph{et al}., CsV$_3$Sb$_5$: a Z2 topological kagome metal with a superconducting ground state. Phys. Rev. Lett. {\bf 125}, 247002 (2020).

\bibitem{BROrtiz_PRM2021_SDWilson} B. R. Ortiz \emph{et al}., Superconductivity in the Z2 kagome metal KV$_3$Sb$_5$. Phys. Rev. Mater. {\bf 5}, 034801 (2021).

\bibitem{YHu_arxiv2021_MShi_VHS} Y. Hu \emph{et al}., Rich Nature of Van Hove Singularities in Kagome Superconductor CsV$_3$Sb$_5$. arxiv:2106.05922 (2021).

\bibitem{MGKang_arxiv2021_RComin} M. G. Kang \emph{et al}., Twofold van Hove singularity and origin of charge order in topological kagome superconductor CsV$_3$Sb$_5$. arxiv:2105.01689 (2021).

\bibitem{ZYHao_2021arxiv_CYChen} Z. Y. Hao \emph{et al}., Dirac Nodal Lines and Nodal Loops in a Topological Kagome Superconductor CsV$_3$Sb$_5$. arxiv:2111.02639 (2021).



\bibitem{YXJiang_NM2020_MZHasan} Y. X. Jiang \emph{et al}., Unconventional chiral charge order in kagome superconductor KV$_3$Sb$_5$. Nat. Mater. {\bf 20}, 1353 (2021).

\bibitem{HLi_arxiv2021_IZeljkovic} H. Li \emph{et al}., Rotation symmetry breaking in the normal state of a kagome superconductor KV$_3$Sb$_5$. arxiv:2104.08209 (2021).

\bibitem{HSXu_PRL2021_DLFeng} H. S. Xu \emph{et al}., Multiband Superconductivity with Sign-Preserving Order Parameter in Kagome Superconductor CsV$_3$Sb$_5$. Phys. Rev. Lett. {\bf 127}, 187004 (2021).

\bibitem{ZHLiu_2021PRX_SCWang} Z. H. Liu \emph{et al}., Charge-Density-Wave-Induced Bands Renormalization and Energy Gaps in a Kagome Superconductor RbV$_3$Sb$_5$. Phys. Rev. X {\bf 11}, 041010 (2021).

\bibitem{NShumiya_PRB2021_MZHasan} N Shumiya \emph{et al}., Intrinsic nature of chiral charge order in the kagome superconductor RbV$_3$Sb$_5$. Phys. Rev. B {\bf 104}, 035131 (2021).

\bibitem{ZWWang_PRB2021_YGYao} Z. W. Wang \emph{et al}., Electronic nature of chiral charge order in the kagome superconductor CsV$_3$Sb$_5$. Phys. Rev. B {\bf 104}, 075148 (2021).

\bibitem{QWang_AM2021_YPQi} Q. Wang \emph{et al}., Charge Density Wave Orders and Enhanced Superconductivity under Pressure in the Kagome Metal CsV$_3$Sb$_5$. Adv. Mater. {\bf 33}, 2102813 (2021).

\bibitem{ZWLiang_PRX2021_XHChen} Z. W. Liang \emph{et al}., Three-dimensional charge density wave and robust zero-bias conductance peak inside the superconducting vortex core of a kagome superconductor CsV$_3$Sb$_5$. Phys. Rev. X {\bf 11}, 031026 (2021).

\bibitem{HXTan_PRL2021_BHYan} H. X. Tan \emph{et al}., Charge density waves and electronic properties of superconducting kagome metals. Phys. Rev. Lett. {\bf 127}, 046401 (2021).

\bibitem{YPLin_PRB2021_RMNandkishore} Y. P. Lin \emph{et al}., Complex charge density waves at Van Hove singularity on hexagonal lattices: Haldane-model phase diagram and potential realization in the kagome metals AV$_3$Sb$_5$ (A=K, Rb, Cs). Phys. Rev. B {\bf 104}, 045122 (2021).

\bibitem{ZGWang_arxiv2021_ZXZhao} Z. G. Wang \emph{et al}., Distinctive momentum dependent charge-density-wave gap observed in CsV$_3$Sb$_5$ superconductor with topological kagome lattice. arXiv:2104.05556 (2021).

\bibitem{BROrtiz_PRX2021_SDWilson_CDW} B. R. Ortiz \emph{et al}., Fermi Surface Mapping and the Nature of Charge-Density-Wave Order in the Kagome Superconductor CsV$_3$Sb$_5$. Phys. Rev. X {\bf 11}, 041030 (2021).

\bibitem{JLuo_arxiv2021_GQZheng} J. Luo \emph{et al}., Star-of-David pattern charge density wave with additional modulation in the kagome superconductor CsV$_3$Sb$_5$ revealed by $^{51}$V-NMR and $^{121/123}$Sb-NQR. arxiv:2108.10263 (2021).

\bibitem{KNakayama_PRB2021_TSato} K. Nakayama \emph{et al}., Multiple energy scales and anisotropic energy gap in the charge-density-wave phase of the kagome superconductor CsV$_3$Sb$_5$. Phys. Rev. B {\bf 104}, L161112 (2021).


\bibitem{YHu_arxiv2021_MShi} Y. Hu \emph{et al}., Charge-order-assisted topological surface states and flat bands in the kagome superconductor CsV$_3$Sb$_5$. arxiv:2104.12725 (2021).


\bibitem{YLuo_arxiv2021_JFHe} Y. Luo \emph{et al}., Distinct band reconstructions in kagome superconductor CsV$_3$Sb$_5$. arXiv:2106.01248 (2021).

\bibitem{RLuo_arxiv2021_SBorisenko} R. Luo \emph{et al}., Charge-Density-Wave-Induced Peak-Dip-Hump Structure and Flat Band in the Kagome Superconductor CsV$_3$Sb$_5$. arxiv:2106.06497 (2021).

\bibitem{HMiao_PRB2021_BHYan} H. Miao \emph{et al}., Geometry of the charge density wave in the kagome metal AV$_3$Sb$_5$. Phys. Rev. B {\bf 104}, 195132 (2021).

\bibitem{HLLuo_arxiv2021_XJZhou} H. L. Luo \emph{et al}., Electronic Nature of Charge Density Wave and Electron-Phonon Coupling in Kagome Superconductor KV$_3$Sb$_5$. arxiv:2107.02688 (2021).



\bibitem{JWYu_arxiv2021_WLi} J. W. Yu \emph{et al}., Evolution of electronic structure in pristine and hole-doped kagome metal RbV$_3$Sb$_5$. arxiv:2109.11286 (2021).





\bibitem{CCZhao_arxiv2021_SYLi} C. C. Zhao \emph{et al}., Nodal superconductivity and superconducting domes in the topological Kagome metal CsV$_3$Sb$_5$. arxiv:2102.08356 (2021).

\bibitem{HChen_Nature2021_HJGao} H. Chen \emph{et al}., Roton pair density wave and unconventional strong-coupling superconductivity in a topological kagome metal. Nature {\bf 599}, 222 (2021).

\bibitem{QWYin_CPL2021_HCLei} Q. W. Yin \emph{et al}., Superconductivity and normal-state properties of kagome metal RbV$_3$Sb$_5$ single crystals. Chin. Phys. Lett. {\bf 38}, 037403 (2021).

\bibitem{BROrtiz_PRM2019_ESToberer} B. R. Ortiz \emph{et al}., New kagome prototype materials: discovery of KV$_3$Sb$_5$, RbV$_3$Sb$_5$, and CsV$_3$Sb$_5$. Phys. Rev. Mater. {\bf 3}, 094407 (2019).


\bibitem{CMielkeIII_arxiv2021_ZGuguchia} C. Mielke III \emph{et al}., Time-reversal symmetry-breaking charge order in a correlated kagome superconductor. arxiv:2106.13443 (2021).

\bibitem{KYChen_PRL2021_JGCheng} K. Y. Chen \emph{et al}., Double Superconducting Dome and Triple Enhancement of T$_c$ in the Kagome Superconductor CsV$_3$Sb$_5$ under high pressure. Phys. Rev. Lett. {\bf 126}, 247001 (2021).

\bibitem{FDu_PRB2021_HYuan} F. Du \emph{et al}., Pressure-induced double superconducting domes and charge instability in the kagome metal KV$_3$Sb$_5$. Phys. Rev. B {\bf 103}, L220504 (2021).

\bibitem{XXWu_PRL2021_RThomale} X. X. Wu \emph{et al}., Nature of Unconventional Pairing in the Kagome Superconductors AV$_3$Sb$_5$ (A=K, Rb, Cs). Phys. Rev. Lett. {\bf 127}, 177001 (2021).




\bibitem{HXLi_PRX2021_HMiao} H. X. Li \emph{et al}., Observation of unconventional charge density wave without acoustic phonon anomaly in kagom Superconductors AV$_3$Sb$_5$ (A=Rb,Cs). Phys.Rev. X, {\bf 11}, 031050 (2021).

\bibitem{DWSong_arxiv2021_XHChen} D. W. Song \emph{et al}., Orbital ordering and fluctuations in a kagome superconductor CsV$_3$Sb$_5$. arxiv:2104.09173 (2021).




\bibitem{XXZhou_2021PRB_HHWen} X. X. Zhou \emph{et al}., Origin of the Charge Density Wave in the Kagome Metal CsV$_3$Sb$_5$ as Revealed by Optical Spectroscopy. Phys. Rev. B {\bf 104}, L041101 (2021).

\bibitem{TPark_2021PRB_LBalents} T. Park \emph{et al}., Electronic instabilities of kagome metals: saddle points and Landau theory. Phys. Rev. B {\bf 104}, 035142 (2021).

\bibitem{YPLin_2021PRB_RMNandkishore} Y. P. Lin \emph{et al}., Complex charge density waves at Van Hove singularity on hexagonal lattices: Haldane-model phase diagram and potential realization in the kagome metals AV$_3$Sb$_5$ (A=K, Rb, Cs). Phys. Rev. B {\bf 104}, 045122 (2021).

\bibitem{HXTan_2021PRL_BHYan} H. X. Tan \emph{et al}., Charge Density Waves and Electronic Properties of Superconducting Kagome Metals. Phys. Rev. Lett. {\bf 127}, 046401 (2021).

\bibitem{NRatcliff_2021PRM_JWHarter} N. Ratcliff \emph{et al}., Coherent phonon spectroscopy and interlayer modulation of charge density wave order in the kagome metal CsV$_3$Sb$_5$. Phys. Rev. Mater. {\bf 5}, L111801 (2021).

\bibitem{EUykur_2021arxiv_AATsirlin} E. Uykur \emph{et al}., Optical detection of charge-density-wave instability in the non-magnetic kagome metal KV$_3$Sb$_5$. arxiv:2103.07912 (2021).



\bibitem{CZWang_arxiv2021_JHCho} C. Z. Wang \emph{et al}., Origin of Charge Density Wave in Layered Kagome Metal CsV$_3$Sb$_5$. arxiv:2109.01921 (2021).


\bibitem{YFXie_2021arxiv_PCDai} Y. F. Xie \emph{et al}., Electron-phonon coupling in the charge density wave state of CsV$_3$Sb$_5$. arxiv:2111.00654 (2021).

\bibitem{YQCai_arxiv2021_CYChen} Y. Q. Cai \emph{et al}., Emergence of Quantum Confinement in Topological Kagome Superconductor CsV$_3$Sb$_5$ family. arxiv:2109.12778 (2021).

\bibitem{HZhao_Nature2021_IZeljkovic} H. Zhao \emph{et al}., Cascade of correlated electron states in a kagome superconductor CsV$_3$Sb$_5$. Nature {\bf 599}, 216 (2021).

\bibitem{YPSong_PRL2021_XLChen} Y. P. Song \emph{et al}., Competition of Superconductivity and Charge Density Wave in Selective Oxidized CsV$_3$Sb$_5$ Thin Flakes. Phys. Rev. Lett. {\bf 127}, 237001 (2021).

\bibitem{BQSong_arxiv2021_SYLi} B. Q. Song \emph{et al}., Competing superconductivity and charge-density wave in Kagome metal CsV$_3$Sb$_5$: evidence from their evolutions with sample thickness. arxiv:2105.09248 (2021).




\end{thebibliography}
\end{document}